%% file: conference_gcram_ecc.tex
\documentclass[conference]{IEEEtran}
\usepackage[OT1]{fontenc}
\IEEEoverridecommandlockouts
\usepackage{cite}
\usepackage{float}
\usepackage{graphicx}
\usepackage{color}
\usepackage{hyperref}
\usepackage{subfigure}
\usepackage{mathrsfs,amsfonts,amssymb,amsmath}
\usepackage{enumerate}
\usepackage{algorithm}
\usepackage{algorithmic}
\usepackage{multirow}
\usepackage{xargs}  
\usepackage[dvipsnames]{xcolor}
\usepackage{blkarray}
\usepackage{setspace}
\usepackage{amsthm}
\usepackage{threeparttable}
\usepackage{colortbl}
\usepackage{booktabs}
\usepackage{makecell}
\usepackage{url} 
\usepackage{tikz}
\usepackage{pgfplots}
\pgfplotsset{compat=1.18}
\usepackage[colorinlistoftodos,prependcaption]{todonotes}
\newcommandx{\info}[2][1=]{\todo[linecolor=OliveGreen,backgroundcolor=OliveGreen!25,bordercolor=OliveGreen,#1]{#2}}

\definecolor{mygray}{gray}{.9}
\definecolor{emerald}{rgb}{0.31, 0.78, 0.47}
\definecolor{eccgrid}{gray}{0.64}
\definecolor{ecctone}{RGB}{140,86,75}
\definecolor{eccttwo}{RGB}{255,127,14}
\definecolor{ecctthree}{RGB}{44,160,44}
\definecolor{ecctfour}{RGB}{214,39,40}
\newcommand{\eccplotfont}{\footnotesize}
\newcommand{\eccleftblockwidth}{0.625\textwidth}
\newcommand{\eccbarcolwidth}{0.365\textwidth}
\newcommand{\eccmaincolsep}{-0.05\textwidth}
\newcommand{\eccsmallcolwidth}{0.495\linewidth}
\newcommand{\eccsmallcolsep}{-0.05\linewidth}
\newcommand{\eccsubfiginnerwidth}{\dimexpr\linewidth-7pt\relax}
\newcommand{\eccaxisheight}{3.4cm}
\newcommand{\eccbaraxisheight}{6.35cm}
\newcommand{\eccsubfigrowsep}{-1mm}
\newcommand{\ecclegendrowsep}{1mm}
\newcommand{\ecctlegendbox}{%
    \begingroup
    \setlength{\fboxsep}{1pt}%
    \fbox{%
        \eccplotfont
        \begin{tabular}{@{\hspace{0.5em}}c@{\hspace{1.3em}}c@{\hspace{1.3em}}c@{\hspace{1.3em}}c@{\hspace{0.5em}}}
        \tikz[baseline=-0.45ex]{\draw[ecctone,line width=0.85pt] (0,0)--(0.38,0); \fill[ecctone] (0.19,0) circle[radius=1.15pt];}~$t=1$ &
        \tikz[baseline=-0.45ex]{\draw[eccttwo,line width=0.85pt] (0,0)--(0.38,0); \fill[eccttwo] (0.15,-0.04) rectangle (0.23,0.04);}~$t=2$ &
        \tikz[baseline=-0.45ex]{\draw[ecctthree,line width=0.85pt] (0,0)--(0.38,0); \fill[ecctthree] (0.19,0.055)--(0.13,-0.045)--(0.25,-0.045)--cycle;}~$t=3$ &
        \tikz[baseline=-0.45ex]{\draw[ecctfour,line width=0.85pt] (0,0)--(0.38,0); \fill[ecctfour] (0.19,0.06)--(0.25,0)--(0.19,-0.06)--(0.13,0)--cycle;}~$t=4$
        \end{tabular}%
    }%
    \endgroup
}
\newcommand{\eccselectionid}[2]{%
    \tikz[baseline=-0.55ex]{%
        \node[circle, draw=black, line width=0.25pt, fill=#1, inner sep=0pt, minimum size=1.18em, font=\eccplotfont] {#2};%
    }%
}
\newcommand{\eccselectionlegendbox}{%
    \begingroup
    \setlength{\fboxsep}{2pt}%
    \fbox{%
        \eccplotfont
        \begin{tabular}{@{\hspace{0.5em}}c@{\hspace{1.2em}}c@{\hspace{1.2em}}c@{\hspace{1.2em}}c@{\hspace{0.5em}}}
        \eccselectionid{ecctone!65}{1}~$\mathrm{ECC}(136,128,1)$ &
        \eccselectionid{eccttwo!78}{2}~$\mathrm{ECC}(274,256,2)$ &
        \eccselectionid{ecctthree!65}{3}~$\mathrm{ECC}(283,256,3)$ &
        \eccselectionid{ecctfour!65}{4}~$\mathrm{ECC}(552,512,4)$
        \end{tabular}%
    }%
    \endgroup
}
\pgfplotsset{
    ecc axis/.style={
        width=\linewidth,
        height=\eccaxisheight,
        tick label style={font=\eccplotfont},
        label style={font=\eccplotfont},
        xlabel style={yshift=2pt},
        ylabel style={xshift=3pt},
        ymajorgrids=true,
        grid style={eccgrid, line width=0.25pt, dashed},
        axis line style={black, line width=0.35pt},
        tick style={black, line width=0.35pt},
        legend style={
            draw=gray!35,
            fill=white,
            fill opacity=0.88,
            text opacity=1,
            font=\eccplotfont,
            cells={anchor=west},
            inner sep=1.4pt
        },
    },
    ecc t1/.style={ecctone, line width=0.85pt, mark=*, mark size=1.45pt},
    ecc t2/.style={eccttwo, line width=0.85pt, mark=square*, mark size=1.45pt},
    ecc t3/.style={ecctthree, line width=0.85pt, mark=triangle*, mark size=1.65pt},
    ecc t4/.style={ecctfour, line width=0.85pt, mark=diamond*, mark size=1.55pt},
    ecc stack0/.style={ybar, fill opacity=0.18, draw=black!55, line width=0.22pt},
    ecc stack1/.style={ybar, fill opacity=0.38, draw=black!55, line width=0.22pt},
    ecc stack2/.style={ybar, fill opacity=0.58, draw=black!55, line width=0.22pt},
    ecc stack3/.style={ybar, fill opacity=0.78, draw=black!55, line width=0.22pt},
    ecc stack4/.style={ybar, fill opacity=1.00, draw=black!55, line width=0.22pt},
}

\def\BibTeX{{\rm B\kern-.05em{\sc i\kern-.025em b}\kern-.08em
    T\kern-.1667em\lower.7ex\hbox{E}\kern-.125emX}}

\begin{document}

\title{Reducing Power Consumption of Embedded Dynamic Memories with ECCs
}

\author{
    \IEEEauthorblockN{Wenqing~Song,
     Yifei~Shen, Andreas~Burg}
    \IEEEauthorblockA{\textit{Telecommunications Circuits Laboratory, EPFL, Lausanne, Switzerland}\\
    {{\{wenqing.song, yifei.shen, andreas.burg\}@epfl.ch}}}

}
\maketitle

\begin{abstract}
  Gain-cell embedded dynamic random-access memory (GCRAM) offers dense and energy-efficient on-chip storage, but retention-time variations force frequent refresh operations to cover worst-case bits. Error-correction codes (ECCs) can alleviate this limitation by masking bit errors from weak cells and thereby reduce refresh cost. However, the trade-off between the additional access and logic energy introduced by ECCs and the power savings from longer refresh intervals is nontrivial, especially considering the wide range of available ECC options. To optimize overall power consumption, we propose an ECC selection method that combines a refresh-interval model with power analysis to identify the minimum-power ECC configurations under a given yield constraint. Across different memory bandwidths, activity factors, and read/write ratios, the evaluation results show that the best ECC option shifts from stronger codes in refresh-dominated operating regions to lower-overhead codes in access-dominated regions and achieves $46.8\%$ to $94.8\%$ reduction in total power relative to the no-ECC reference.
\end{abstract}


\section{Introduction}\label{sec:intro}
{
{On-chip memories are a major contributor to the area and energy of signal-processing and artificial-intelligence~(AI) systems.} Among various memory technologies, gain-cell embedded dynamic random-access memory~(GCRAM) provides a dense and energy-efficient alternative to $6$T SRAM because gain-cell bit cells use fewer transistors~\cite{chun2011667,chun20113t}. {Recent studies further motivate the use of GCRAM in accelerator memory hierarchies with diverse bandwidth and data-lifetime requirements~\cite{li2025gainsight,wang2025opengcram}}. The main challenge of GCRAM is maintaining stored data integrity. Leakage variations create a tail of weak cells with short retention times, which can require frequent refresh even during idle times. As technology scales and leakage sensitivity increases, retention times become shorter~\cite{nguyen2024mcaimem}. In some workloads or architectures (e.g., low-activity scenarios), refresh power of GCRAM system memories can become a dominant component of total memory power because access activity, and therefore access power, is low.

{Error-correction codes~(ECCs) protect memory systems by adding parity bits to each data word, which enables error
detection and correction before the data is used~\cite{choi2018decoder,rohman2024fast}. In retention-limited dynamic memories, ECCs can also relax the refresh requirement~\cite{wilkerson2010reducing}. Instead of setting the refresh interval based on the weakest cells, ECCs allow each codeword to tolerate a limited number of retention errors. Stronger ECCs can support a longer refresh interval under the same target yield, but require additional parity-bit storage, encoding/decoding energy, and latency. The important question is how to choose the ECC scheme that balances refresh interval extension against these overheads in a retention-limited GCRAM system.}

{
Prior work mainly targets off-chip DRAM and high-bandwidth memory~(HBM) systems, where the latency and energy costs of stronger ECCs can be comparatively small relative to off-chip data transfers~\cite{lee2025shift}. Stealth ECC~\cite{lee2022stealth} and CARE~\cite{chen2021care} use stronger or adaptive correction mechanisms to complement conventional single-error correction and double-error detection~(SEC-DED) codes. 
These methods do not directly apply to on-chip memories (especially not GCRAM) where the error behavior, access energy, and timing constraints are very different. Recent ECC studies for on-chip memories mainly target soft errors in SRAMs~\cite{rohman2024fast,joshi2026modeling}, rather than retention errors in GCRAM. Hi-ECC~\cite{wilkerson2010reducing} uses strong ECC with $5$-bit error-correction capability to identify failure-prone cache sections offline and then runs SEC-DED in normal mode to reduce eDRAM cache refresh power. Thus, Hi-ECC does not explore the trade-offs associated with different ECCs for runtime error correction.}

In this paper, we study joint ECC-strength and refresh-interval
selection for retention-limited GCRAM.
The main contributions are as follows:
\begin{itemize}
    \item \textbf{Refresh-yield modeling:}
    We derive a row-repair-aware yield model that converts a calibrated GCRAM retention-time distribution into the maximum feasible refresh interval for each ECC candidate.

    \item \textbf{Activity-aware power optimization:}
    We build an average-power model that jointly accounts for refresh power, GCRAM access energy, parity-bit overhead, and ECC encoder/decoder energy.
    The model selects the minimum-power ECC configuration under workload and chip-yield constraints.

    \item \textbf{BCH-based GCRAM case study:}
    We synthesize BCH encoder and decoder implementations with one- to four-bit correction in $16\,\mathrm{nm}$ technology and use their measured energy and delay in a $16\,\mathrm{MB}$
    GCRAM evaluation.
    The results quantify the transition from stronger ECCs in refresh-dominated regimes to lower-overhead ECCs in access-dominated regimes.
\end{itemize}

The paper is organized as follows. Section~\ref{sec:background} provides background on GCRAM retention errors and ECC basics. Section~\ref{sec:framework} describes the proposed ECC selection method based on power modeling. The ECC evaluation results are presented under different workload conditions in Section~\ref{sec:evaluation}. Section~\ref{sec:conclusion} concludes the paper.

}

\begin{figure}[t]
    \centering
    \includegraphics[width=0.47\textwidth]{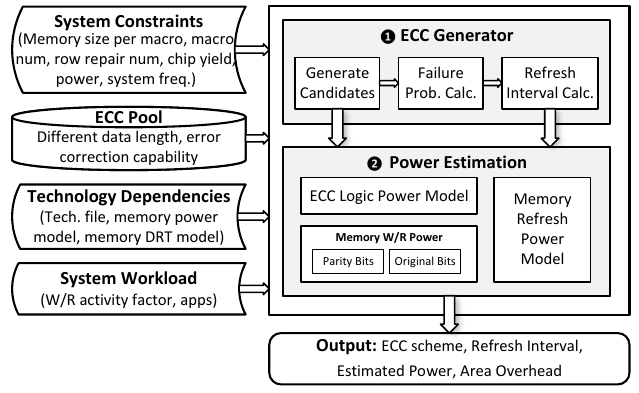}
    \caption{Activity-aware ECC selection flow for GCRAM.}
    \label{fig:framework}
\end{figure}
\section{Background} \label{sec:background}
\subsection{Retention-Limited GCRAM}
GCRAM can be implemented with different device technologies that offer different retention ranges.
Silicon-based~(Si)-GCRAM is attractive for dense on-chip memory in standard CMOS, but its retention time is often limited to the $\mu$s range~\cite{harel202416}. 
Oxide-semiconductor based GCRAM can extend retention to much longer times, in some cases to the order of seconds, although the effective retention still depends on operating conditions and write activity~\cite{li2025gainsight}. 
This work focuses on Si-GCRAM, but the proposed method can be adapted to other GCRAM technologies by recalibrating
the underlying retention model.

To model GCRAM retention errors, we use a data-retention-time~(DRT) distribution rather than a single worst-case retention time. Let $T_{\mathrm{ret}}$ denote the retention time of a GCRAM cell.
For Si-GCRAM cells, $T_{\mathrm{ret}}$ is modeled as a log-normal random variable $T_{\mathrm{ret}} \sim \mathrm{LogNormal}(\mu, \sigma^2)$~\cite{meinerzhagen2018gain}, where $\mu$ and $\sigma$ denote the mean and standard deviation of $\ln(T_{\mathrm{ret}})$, respectively. The parameters $\mu$ and $\sigma$ are technology-specific and can be calibrated from measured or published data. For a refresh interval~$\tau$, a cell is considered failed if $T_{\mathrm{ret}} < \tau$. Therefore, the pre-ECC bit error rate~(pre-BER) at refresh interval~$\tau$ is given by
\begin{equation}
    p_{\tau}
    =
    \Pr\!\left(
        T_{\mathrm{ret}} < \tau
    \right)
    =
    \frac{1}{2}
    \left[
    1 +
    \operatorname{erf}
    \left(
    \frac{\ln(\tau)-\mu}
         {\sigma\sqrt{2}}
    \right)
    \right],
    \label{eq:pre_ecc_ber}
\end{equation}
where $\operatorname{erf}(\cdot)$ denotes the error function. Once the DRT distribution is calibrated, different refresh intervals, memory capacities, and ECC configurations can be evaluated using the same underlying retention model.

\subsection{ECC for Refresh Relaxation}
In this work, we use various binary BCH codes as the ECC pool because they provide multi-bit correction capability with moderate implementation complexity. A BCH code is denoted by $\mathrm{ECC}(n, k, t)$, where $n$ is the codeword length, $k$ is the number of protected data bits, and $t$ represents the maximum number of correctable bit errors per codeword.
For memory integration, $k$ is typically chosen as a power of two to align with common memory row sizes. 
To achieve error correction, the ECC module encodes incoming data and stores it together with parity bits in the memory. Upon a read request, the stored codeword is decoded and up to $t$ errors are corrected using the stored parity bits. 

\begin{figure}[t]
    \centering
    \includegraphics[width=0.453\textwidth]{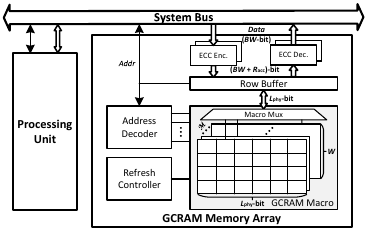}
    \caption{System architecture of an ECC-protected GCRAM module.}
    \label{fig:system}
\end{figure}

\section{Activity-Aware ECC Selection Method} \label{sec:framework}
This section describes the analytical ECC selection method shown in Fig.~\ref{fig:framework}.
The method takes as inputs the memory organization, the activity factors $\alpha_{\mathrm{read}}$ and $\alpha_{\mathrm{write}}$, and the target chip yield $Y_{\mathrm{target}}$.
Memory repair mechanisms are also considered by including the row-repair\footnote{We assume row-repair since it is more easily integrated in the controller, but the selection strategy and results are also valid for bit-/column-repair.} budget~$u$ in the yield model.
The flow first enumerates the ECC candidate set.
For each candidate, the method uses the DRT model and the row-repair budget to determine the largest refresh interval that satisfies the target chip yield.
Then the corresponding refresh power is computed and the memory-access and ECC-logic energies are converted into average power under the specified workload.
The selected candidate is the one with the lowest modeled total power among the candidates.
The GCRAM DRT model, refresh-power data, and access-energy data are calibrated from silicon measurements.
The ECC-logic energy is obtained from synthesized BCH encoder and decoder blocks.
The chip-yield and total-power estimates are computed analytically from these calibrated inputs.

\subsection{System Model}

We model the memory subsystem in Fig.~\ref{fig:system} as $\mathcal{M}$~GCRAM macros connected to a processor or accelerator through a data bus of width $BW$~bits/cycle. Each macro contains $W$~rows, and each logical row stores $K$~logical (data) bits before ECC. For an ECC candidate $c=\mathrm{ECC}(n,k,t)$, the number of ECC codewords per row is
\begin{equation}
    N_{\mathrm{cw}}(c)=\left\lceil\frac{K}{k}\right\rceil .
\end{equation}
The corresponding physical length is
\begin{equation}
    L_{\mathrm{phy}}(c)=N_{\mathrm{cw}}(c)\cdot n,
\end{equation}
which includes both data and parity bits. For $BW$-bit data access, the number of active ECC blocks and the accessed parity bits are
\begin{equation}
    N_{\mathrm{ecc}}(c)=\left\lceil\frac{BW}{k}\right\rceil,\qquad
    R_{\mathrm{acc}}(c)=N_{\mathrm{ecc}}(c)(n-k).
\end{equation}
The no-ECC reference is denoted by $c=0$. For this case, no parity bits are needed, so $L_{\mathrm{phy}}(0)=K$, $R_{\mathrm{acc}}(0)=0$.

Written logical bits are encoded before storage, and read physical bits are decoded before the data are used. A refresh controller periodically triggers raw row-by-row refresh operations at interval~$\tau$. In this work, refresh operations read and rewrite raw rows without going through the ECC process. Cells with~\(T_{\mathrm{ret}}<\tau\) are modeled as failed bits, and correction is applied only when the corresponding word is read by the system. Therefore, ECC logic energy is charged only to memory read/write accesses, not to refresh operations. The ECC logic can also be shared across macros.

\subsection{Yield-Constrained Refresh--Interval Selection} \label{subsec:ecc}

For a refresh interval~$\tau$, the pre-ECC BER $p_{\tau}$ is obtained from the calibrated DRT model in~(\ref{eq:pre_ecc_ber}). Without ECC, the probability that a logical row contains at least one retention error is
\begin{equation}
    \mathcal{P}_{\mathrm{row}}^{(0)}(\tau)=1-(1-p_{\tau})^K .
\end{equation}

With ECC candidate $c=\mathrm{ECC}(n,k,t)$, the probability that one codeword contains exactly $i$ raw-bit errors is
\begin{equation}
    \pi_i(\tau;n)=\binom{n}{i}p_{\tau}^{i}(1-p_{\tau})^{n-i}.
\end{equation}
The probability that all codewords in a row remain correctable is therefore $\left(\sum_{i=0}^{t}\pi_i(\tau;n)\right)^{N_{\mathrm{cw}}(c)}$, and the post-ECC row failure probability is
\begin{equation}
    \mathcal{P}_{\mathrm{row}}^{(c)}(\tau)
    =
    1-
    \left(
    \sum_{i=0}^{t}\pi_i(\tau;n)
    \right)^{N_{\mathrm{cw}}(c)} .
\end{equation}

We include row-repair as a representative repair mechanism. If up to $u$ rows can be repaired per macro, the retention-limited macro yield is
\begin{equation}
    Y_{\mathrm{macro}}(\tau,c,u)
    =
    \sum_{j=0}^{u}
    \binom{W}{j}
    \mathcal{P}_{\mathrm{row}}(\tau,c)^j
    \left(1-\mathcal{P}_{\mathrm{row}}(\tau,c)\right)^{W-j},
\end{equation}
where $\mathcal{P}_{\mathrm{row}}(\tau,c) = \mathcal{P}_{\mathrm{row}}^{(0)}(\tau)$ for no ECC and $\mathcal{P}_{\mathrm{row}}(\tau,c)=\mathcal{P}_{\mathrm{row}}^{(c)}(\tau)$ otherwise. The chip-yield over $\mathcal{M}$ macros is
\begin{equation}
    Y_{\mathrm{chip}}(\tau,c,u)=Y_{\mathrm{macro}}(\tau,c,u)^{\mathcal{M}} .
\end{equation}

For each ECC candidate, the selected refresh interval is the largest interval that satisfies the chip-yield target:
\begin{equation}
    \tau_{\max}(c,u)
    =
    \max_{\tau}
    \left\{
    \tau:
    Y_{\mathrm{chip}}(\tau,c,u)\geq Y_{\mathrm{target}}
    \right\}.
    \label{eq:taumax}
\end{equation}

\subsection{Power Model}

The considered total power consists of refresh power, memory access power derived from per-bit access energies, and ECC logic power derived from per-operation logic energies. Leakage is not included in the selection objective because it is relatively small compared with the dynamic terms and weakly dependent on the ECC candidate.

The \emph{refresh power} is scaled from a calibrated per-macro reference value obtained from silicon. Let $P_{\mathrm{ref},0}$ be the refresh power of one macro at reference interval $\tau_0$. For candidate $c$, the refresh power at $\tau_{\max}(c,u)$ is
\begin{equation}
    P_{\mathrm{refresh}}(c)
    =
    \mathcal{M}P_{\mathrm{ref},0}
    \frac{L_{\mathrm{phy}}(c)}{K}
    \frac{\tau_0}{\tau_{\max}(c,u)}.
\end{equation}
The factor $L_{\mathrm{phy}}(c)/K$ accounts for the additional parity bits that are stored in GCRAM and are refreshed together with the data bits. 

Let $E_{\mathrm{read}}$ and $E_{\mathrm{write}}$ denote the per-bit read and write energies of a stored bit. Since data and parity bits are stored in the same GCRAM array, they use the same per-bit access energies. With operating frequency $f$ and activity factors $\alpha_{\mathrm{read}}$ and $\alpha_{\mathrm{write}}$, the \emph{average memory access power} is
\begin{equation}
\label{eq:p_access}
\begin{aligned}
    P_{\mathrm{access}}(c)
    =&
    f\left(BW+R_{\mathrm{acc}}(c)\right)
    \\
    &\times
    \left(
    \alpha_{\mathrm{read}}E_{\mathrm{read}}
    +
    \alpha_{\mathrm{write}}E_{\mathrm{write}}
    \right).
\end{aligned}
\end{equation}

ECC logic energy is characterized from synthesized encoder and decoder blocks. Let $E_{\mathrm{enc}}(c)$ denote the energy of one encoder operation for candidate $c$, and let $E_{\mathrm{dec}}^{(i)}(c)$ denote the decoder energy for one codeword when $i$ errors are injected. 

Since uncorrectable patterns are handled by the yield and repair model, the decoder energy is averaged over the error distribution of accepted chips whose retention-error count is within the ECC correction capability. 

Defining $\tau_c=\tau_{\max}(c,u)$ and $S_c(\tau_c)=\sum_{i=0}^{t}\pi_i(\tau_c;n)$, the average decoder energy per codeword is
\begin{equation}
\begin{aligned}
    \overline{E}_{\mathrm{dec}}(c)
    =
    \sum_{i=0}^{t}
    \frac{\pi_i(\tau_c;n)}{S_c(\tau_c)}
    E_{\mathrm{dec}}^{(i)}(c).
\end{aligned}
\end{equation}

The \emph{ECC logic power} for one access path at frequency~$f$ is then
\begin{equation}
\label{eq:p_ecc}
    P_{\mathrm{ecc}}(c)
    =
    f\,N_{\mathrm{ecc}}(c)
    \left(
    \alpha_{\mathrm{write}}E_{\mathrm{enc}}(c)
    +
    \alpha_{\mathrm{read}}\overline{E}_{\mathrm{dec}}(c)
    \right).
\end{equation}
The \emph{total memory subsystem power for candidate $c$} is
\begin{equation}
    P_{\mathrm{total}}(c)
    =
    P_{\mathrm{refresh}}(c)
    +
    P_{\mathrm{access}}(c)
    +
    P_{\mathrm{ecc}}(c).
\end{equation}
The selected ECC and refresh interval are therefore
\begin{equation}
    c^{\star}
    =
    \arg\min_{c\in\mathcal{C}}
    P_{\mathrm{total}}(c),
    \qquad
    \tau^{\star}=\tau_{\max}(c^{\star},u),
\end{equation}
where $\mathcal{C}$ contains no-ECC and all ECC candidates satisfying any potential maximum storage-overhead constraint.
\begin{figure*}[t]
    \centering
    \begin{minipage}[T]{\eccleftblockwidth}
        \centering
        \ecctlegendbox
        \\[\ecclegendrowsep]
        \begingroup
        \begin{minipage}[T]{\eccsmallcolwidth}
            \centering
            \subfigure[Parity-bit overhead.]{
                \begin{minipage}[T]{\eccsubfiginnerwidth}
                \centering
                \input{figure/ecc_parity_overhead_plot.tex}
                \end{minipage}
                \label{fig:ecc_storage_overhead}
            }
        \end{minipage}%
        \hspace{\eccsmallcolsep}%
        \begin{minipage}[T]{\eccsmallcolwidth}
            \centering
            \setcounter{subfigure}{2}%
            \subfigure[Encoder energy.]{
                \begin{minipage}[T]{\eccsubfiginnerwidth}
                \centering
                \input{figure/ecc_encoder_power_plot.tex}
                \end{minipage}
                \label{fig:ecc_encoder_energy}
            }
        \end{minipage}
        \\[\eccsubfigrowsep]
        \begin{minipage}[T]{\eccsmallcolwidth}
            \centering
            \setcounter{subfigure}{1}%
            \subfigure[Decoder delay.]{
                \begin{minipage}[T]{\eccsubfiginnerwidth}
                \centering
                \input{figure/ecc_decoder_delay_plot.tex}
                \end{minipage}
                \label{fig:ecc_decoder_delay}
            }
        \end{minipage}%
        \hspace{\eccsmallcolsep}%
        \begin{minipage}[T]{\eccsmallcolwidth}
            \centering
            \setcounter{subfigure}{3}%
            \subfigure[Encoder delay.]{
                \begin{minipage}[T]{\eccsubfiginnerwidth}
                \centering
                \input{figure/ecc_encoder_delay_plot.tex}
                \end{minipage}
                \label{fig:ecc_encoder_delay}
            }
        \end{minipage}
        \endgroup
    \end{minipage}%
    \hspace{\eccmaincolsep}%
    \begin{minipage}[T]{\eccbarcolwidth}
        \centering
        \subfigure[Decoder energy.]{
            \begin{minipage}{\eccsubfiginnerwidth}
            \centering
            \input{figure/ecc_decoder_power_plot.tex}
            \end{minipage}
            \label{fig:ecc_decoder_energy}
        }
    \end{minipage}
    \vspace{-2mm}
    \caption{BCH implementation cost at $16$\,nm. }
    \label{fig:ecc_cost}
    \vspace{-3mm}
\end{figure*}
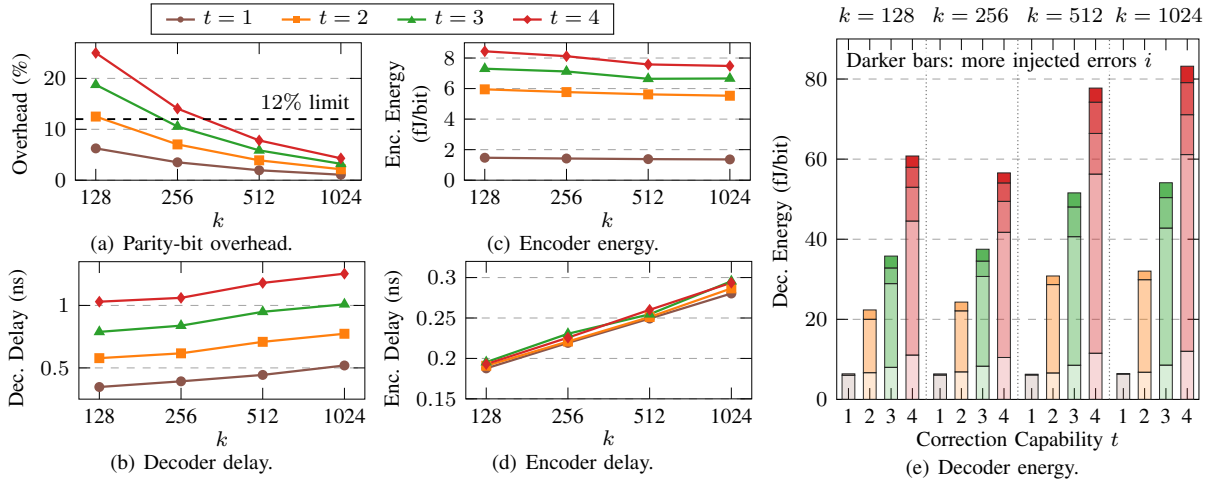

\section{Refresh--ECC Trade-off Evaluation} \label{sec:evaluation}
\subsection{Evaluation Setup and ECC Cost Characterization}
We evaluate the trade-offs based on the model in Sec.~\ref{sec:framework} for a $16$\,nm FinFET technology. For our case study, we consider a GCRAM memory with  $16$\,MB of logical data, organized as $\mathcal{M}=256$ macros. Each macro has $W=512$ rows, and each logical row stores $K=1024$ data bits before ECC. Unless stated otherwise, the target chip yield is $Y_{\mathrm{target}}=90\%$, the row-repair budget is $u=5$ repairable rows per macro, and the operating frequency is $f=500$\,MHz.

The GCRAM refresh-power, access-energy, and DRT models are calibrated from silicon measurements reported in~\cite{andac25a512}. At the reference interval $\tau_0=100\,\mu$s, the refresh power of one macro is $P_{\mathrm{ref},0}=112\,\mu$W. The read and write energies are $E_{\mathrm{read}}=20.8$\,fJ/bit and $E_{\mathrm{write}}=12.8$\,fJ/bit, respectively. The ECC candidate set $\mathcal{C}$ contains the no-ECC reference and BCH codes with $k\in\{128,256,512,1024\}$ and $t\in\{1,2,3,4\}$. ECCs with parity-storage overhead above $12\%$ are excluded from the selection, as shown in Fig.~\ref{fig:ecc_storage_overhead}.

The BCH encoders and decoders are synthesized in the same $16$\,nm technology, and the energy and delay results are presented in Fig.~\ref{fig:ecc_cost}. The encoder is implemented with $\mathrm{GF}(2)$ matrix-vector multiplication, and the decoder is implemented with a lookup-table~(LUT)-based syndrome search for $t=1$ and with the Peterson-based fully parallel decoder in~\cite{fougstedt2019energy} for $t>1$. The delay is measured from input to output. The encoder and decoder energies are obtained from post-synthesis gate-level simulations by dividing active-operation dynamic power by the operating frequency. The resulting energy values are normalized to fJ/bit across codes for plotting, while the power model uses the corresponding full-block operation energies. For the evaluated code lengths, the encoder delay is below $0.3$\,ns (Fig.~\ref{fig:ecc_encoder_delay}) and the decoder delay is below $1.3$\,ns (Fig.~\ref{fig:ecc_decoder_delay}). These delays are compatible with the $500$\,MHz evaluation frequency and with GCRAM operation up to $800$\,MHz in a pipelined implementation. 

Since the encoder only consists of linear operations, the encoder energy per bit in Fig.~\ref{fig:ecc_encoder_energy} is relatively insensitive to~$k$, and increases with $t$ because stronger ECCs generate more parity bits. The decoder energy in Fig.~\ref{fig:ecc_decoder_energy} increases with $k$ and $t$, but the dependence on the number of injected errors~$i$ is not linear. The zero-error case only activates the syndrome-computation logic and therefore has the lowest switching activity. For $t>1$, a nonzero syndrome activates additional logic for error-location and correction. Further injected errors change the internal switching pattern, but they do not activate an independent additional decoding path for each error.
Therefore, the energy increase from $i=0$ to $i=1$ can be larger than the increase between higher error counts.

\subsection{Refresh Relaxation and Yield}
\begin{figure}[t]
    \centering
    \input{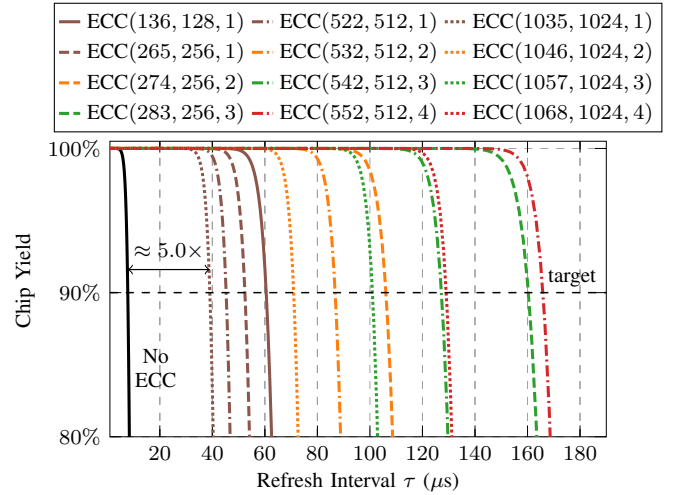}
    \vspace{-8mm}
    \caption{Post-ECC chip yield versus refresh interval of a $16$\,MB memory under different ECC schemes.}
    \label{fig:post_ecc_yield}
    \vspace{-4mm}
\end{figure}
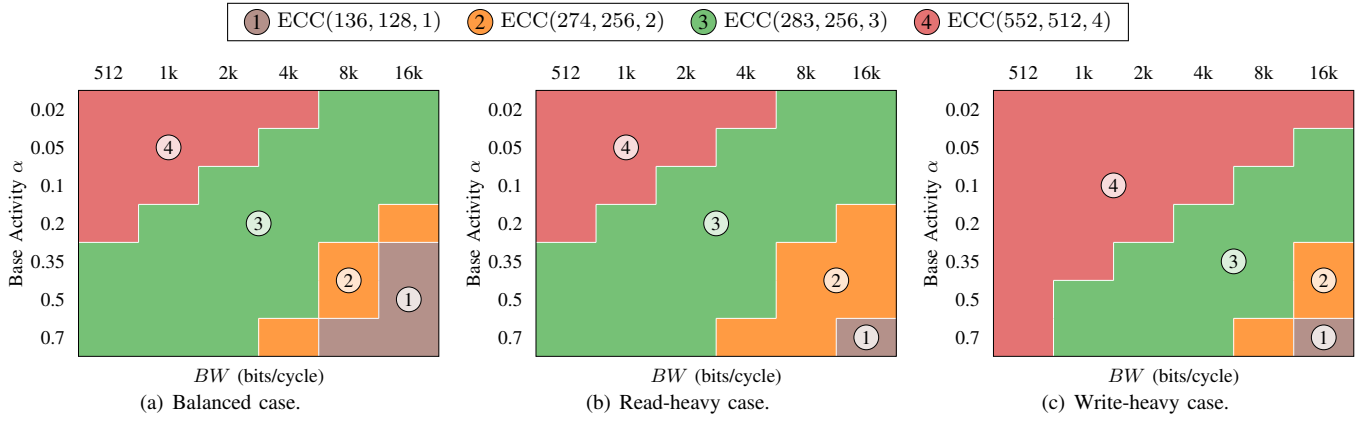
\begin{figure*}[!t]
    \centering
    \eccselectionlegendbox
    \\[0.5mm]
    \begingroup
    \setlength{\subfigcapskip}{-4mm}
    \setlength{\subfigbottomskip}{0pt}
    \subfigure[Balanced case.]{
        \begin{minipage}[T]{0.319\textwidth}
        \centering
        \input{figure/workload_selection_map_balanced.tex}
        \end{minipage}
        \label{fig:selection_map1}
    }%
    \hfill
    \subfigure[Read-heavy case.]{
        \begin{minipage}[T]{0.319\textwidth}
        \centering
        \input{figure/workload_selection_map_read_heavy.tex}
        \end{minipage}
        \label{fig:selection_map2}
    }%
    \hfill
    \subfigure[Write-heavy case.]{
        \begin{minipage}[T]{0.319\textwidth}
        \centering
        \input{figure/workload_selection_map_write_heavy.tex}
        \end{minipage}
        \label{fig:selection_map3}
    }
    \endgroup
    \caption{ECC selections under different workload conditions. }
    \label{fig:selection_map}
    \vspace{-3mm}
\end{figure*}

The refresh interval $\tau_{\max}(c,u)$ is determined by the yield constraint before activity-dependent power is considered. Fig.~\ref{fig:post_ecc_yield} shows $Y_{\mathrm{chip}}(\tau,c,u)$ for a $16$\,MB memory, the row-repair budget $u=5$, and the target chip yield $Y_{\mathrm{target}}=90\%$. The number of ECC codewords per row $N_{\mathrm{cw}}(c)$ is $8$ for ECC with $k=128$, $4$ for ECC with $k=256$, $2$ for ECC with $k=512$, and $1$ for ECC with $k=1024$.

For the fixed macro count and row-repair budget, the chip-yield curves cross the $90\%$ target at the same post-ECC row-failure probability of approximately $1.83\times10^{-3}$. However, ECC changes the pre-ECC bit-error rate that can be tolerated at this row-failure probability. The no-ECC design achieves yield at $p_{\tau}=1.79\times10^{-6}$, whereas the strongest $\mathrm{ECC}(552,512,4)$ tolerates $p_{\tau}=1.32\times10^{-3}$.

Without ECC, the largest refresh interval satisfying the $90\%$ yield target is $7.80\,\mu$s. The lowest parity-storage overhead (only $1.07\%$) code, $\mathrm{ECC}(1035,1024,1)$, extends the refresh interval to $39.05\,\mu$s, which is $5.0\times$ longer than the no-ECC reference. Stronger codes further shift the refresh interval. The candidate with the largest feasible refresh interval, $\mathrm{ECC}(552,512,4)$, prolongs the refresh interval to $165.87\,\mu$s, which provides a $21.3\times$ longer interval than the no-ECC reference with a parity-storage overhead of $7.81\%$. These results confirm that ECC can significantly relax the refresh requirement of GCRAM and reduce refresh power. The remaining question is whether this reduction outweighs the additional parity-access and ECC-logic costs in overall memory-subsystem power.

\subsection{ECC Selection Across Workloads}
To evaluate the impact of different workload conditions on ECC selection, we sweep the effective access width $BW$ from $512$ to $16384$~bits/cycle and the base activity level $\alpha$ over $\{0.02,0.05,0.10,0.20,0.35,0.50,0.70\}$\footnote{These values are typically obtained through multiple memory ports or simultaneous accesses to multiple macros.}. Three different read/write activity ratios are considered. The balanced case sets $\alpha_{\mathrm{read}}=\alpha_{\mathrm{write}}=\alpha$. The read-heavy case sets $\alpha_{\mathrm{read}}=\alpha$ and $\alpha_{\mathrm{write}}=0.25\alpha$. The write-heavy case sets $\alpha_{\mathrm{read}}=0.25\alpha$ and $\alpha_{\mathrm{write}}=\alpha$.
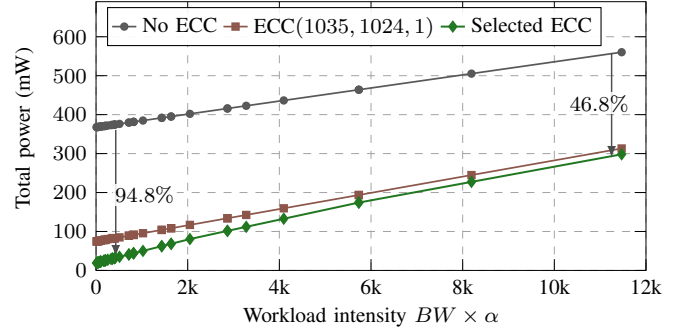
\begin{figure}[t]
    \centering
    \input{figure/workload_power_sweep_balanced.tex}
    \vspace{-5mm}
    \caption{Total power across different workload points in the balanced case. }
    \label{fig:sweep_power}
    \vspace{-7mm}
\end{figure}

Fig.~\ref{fig:selection_map} shows the ECC candidate that minimizes $P_{\mathrm{total}}(c)$ at each workload point. 
We observe that the best ECC option is not always the one with the longest refresh interval. 
At low~$BW$ and low~$\alpha$, $\mathrm{ECC}(552,512,4)$ is selected because refresh power dominates the total power. In this region, the refresh-power saving from stronger ECC is larger than the additional parity-access and decoder-energy costs.
As $BW$ or $\alpha$ increases, the activity-dependent terms in~(\ref{eq:p_access}) and~(\ref{eq:p_ecc}) become more
important. The selected ECC then moves toward lower-overhead codes, first to $\mathrm{ECC}(283,256,3)$ and then to $\mathrm{ECC}(274,256,2)$ or $\mathrm{ECC}(136,128,1)$ in the highest-activity region. 
The read/write mix changes the transition because read and write accesses activate different ECC components.
Read accesses activate the decoder, whose energy overhead is larger for stronger codes.
Write accesses activate the encoder, which has a smaller and nearly constant per-bit energy overhead.
Therefore, reducing the read activity makes stronger correction more attractive.
In the balanced and read-heavy cases, $\mathrm{ECC}(552,512,4)$ is selected in $10$ of the $42$ workload points.
In the write-heavy case, the same code is selected in $22$ of the $42$ workload points because the decoder-energy penalty is reduced.
Conversely, $\mathrm{ECC}(136,128,1)$ is selected in $4$ high-activity points in the balanced case, but only at the highest $BW$ and $\alpha$ in the read-heavy and write-heavy cases.
These results show that ECC selection depends on both the total access activity and the read/write composition of that activity.

Fig.~\ref{fig:sweep_power} compares the total power of the selected ECC, the no-ECC reference, and the lowest-overhead $\mathrm{ECC}(1035,1024,1)$, for the balanced-case workload sweep. The selected ECC is the candidate with the lowest $P_{\mathrm{total}}(c)$ for each workload point. Relative to the no-ECC reference, the selected ECC reduces $P_{\mathrm{total}}(c)$ by $46.8\%$ to $94.8\%$. 
The largest reduction occurs at $BW=512$~bits/cycle and $\alpha=0.02$, where the no-ECC reference is dominated by refresh power. The selected $\mathrm{ECC}(552,512,4)$ reduces the total power from $367.80$\,mW to $19.05$\,mW, and also reduces the total power by $74.4\%$ compared with $\mathrm{ECC}(1035,1024,1)$. 
At the highest workload point, $BW=16384$~bits/cycle and $\alpha=0.70$, the selected $\mathrm{ECC}(136,128,1)$ reduces the total power from $560.30$\,mW to $298.14$\,mW, but the gain over $\mathrm{ECC}(1035,1024,1)$ shrinks to $4.8\%$. The reduced margin indicates that stronger correction becomes less beneficial when access activity makes ECC logic and parity-access overheads a significant part of the total-power objective.

\subsection{Discussion}
The results highlight a separation between reliability-driven and workload-driven design choices.
For a given memory organization and repair budget, the DRT and yield models determine the feasible refresh interval of each ECC candidate independently of the workload.
The workload mainly determines how much of the ECC overhead is paid during normal operation through additional parity movement and encoder or decoder activity.
Therefore, the strongest code is not necessarily the best system-level choice, even when it provides the longest refresh interval.
The benefit of ECC depends on whether the saved refresh power is large enough to compensate for the extra active-energy cost.
This observation suggests that ECC selection for GCRAM should be guided by the relative weight of idle refresh power and active access power, rather than by error correction capability alone.

Memory capacity introduces an additional scaling effect.
For the same chip-yield target, a larger memory exposes more cells and rows to retention failures, which tightens the allowable pre-ECC BER.
At the same time, the aggregate refresh power increases with capacity, whereas the ECC logic energy of a fixed-width access interface is paid per access and does not scale directly with the total memory size.

This difference in scaling shifts the selected ECC strength as capacity increases.
At the moderate-bandwidth, moderate-activity anchor, $\mathrm{ECC}(283,256,3)$ is selected for $8\,\mathrm{MB}$ and $16\,\mathrm{MB}$ memories, whereas $\mathrm{ECC}(552,512,4)$ is selected from $32\,\mathrm{MB}$ onward. The larger memories therefore justify the stronger decoder because the refresh-power reduction becomes more important.
At the high-bandwidth anchor, the selected code changes from $\mathrm{ECC}(136,128,1)$ at $8\,\mathrm{MB}$ to $\mathrm{ECC}(274,256,2)$ at $16\,\mathrm{MB}$ and $\mathrm{ECC}(283,256,3)$ at $32\,\mathrm{MB}$ and above. This trend shows that increasing capacity favors stronger correction, while high access activity still limits the benefit of the strongest code because its decoder and parity-access overheads become visible in the total-power objective.

\section{Conclusion} \label{sec:conclusion}
In this paper, we present an activity-aware ECC selection method for GCRAM that jointly selects refresh interval and ECC strength under chip-yield constraints. By combining post-ECC yield analysis with average-power modeling, our proposed method identifies the ECC configuration that minimizes the total power for a given memory organization and workload activity. Using a $16$\,MB GCRAM memory system as a case study, we show that ECC extends the feasible refresh interval and reduces total power by $46.8\%$ to $94.8\%$ across the swept workloads relative to the no-ECC reference. The selected correction strength depends strongly on the activity and bandwidth. Stronger ECCs are favored when refresh dominates, whereas lower-overhead ECCs are favored when the access and ECC logic terms dominate the activity-dependent power. Beyond GCRAM, the methodology can be adapted to other retention-limited memories such as emerging non-volatile memories and conventional DRAM.

\section*{Acknowledgment}
This work was supported by the Swiss State Secretariat for Education, Research, and Innovation (SERI) under the SwissChips initiative. 

\vfill
\footnotesize
\bibliographystyle{IEEEtran}
\bibliography{IEEEabrv,mybib}

\end{document}

%% file: figure/ecc_parity_overhead_plot.tex
\begin{tikzpicture}
\begin{axis}[
    ecc axis,
    xmin=0.75, xmax=4.25,
    ymin=0, ymax=27,
    xtick={1,2,3,4},
    xticklabels={128,256,512,1024},
    xlabel={$k$},
    ylabel={Overhead (\%)},
    xlabel style={yshift=2pt},
    ylabel style={xshift=-4pt, yshift=-2pt},
]
\addplot[ecc t1] coordinates {(1,6.250) (2,3.516) (3,1.953) (4,1.074)};
\addplot[ecc t2] coordinates {(1,12.500) (2,7.031) (3,3.906) (4,2.148)};
\addplot[ecc t3] coordinates {(1,18.750) (2,10.547) (3,5.859) (4,3.223)};
\addplot[ecc t4] coordinates {(1,25.000) (2,14.062) (3,7.812) (4,4.297)};
\addplot[black, dashed, line width=0.7pt, forget plot] coordinates {(0.75,12) (4.25,12)};
\node[anchor=south east, font=\eccplotfont] at (axis cs:4.22,12) {$12\%$ limit};
\end{axis}
\end{tikzpicture}

%% file: figure/ecc_encoder_power_plot.tex
\begin{tikzpicture}
\begin{axis}[
    ecc axis,
    xmin=0.75, xmax=4.25,
    ymin=0, ymax=9,
    xtick={1,2,3,4},
    xticklabels={128,256,512,1024},
    xlabel={$k$},
    ylabel={\shortstack{Enc. Energy\\(fJ/bit)}},
    xlabel style={yshift=2pt},
    ylabel style={xshift=-4pt, yshift=-2pt},
]
\addplot[ecc t1] coordinates {(1,1.47) (2,1.42) (3,1.38) (4,1.36)};
\addplot[ecc t2] coordinates {(1,5.95) (2,5.77) (3,5.62) (4,5.53)};
\addplot[ecc t3] coordinates {(1,7.30) (2,7.12) (3,6.64) (4,6.66)};
\addplot[ecc t4] coordinates {(1,8.44) (2,8.12) (3,7.58) (4,7.48)};
\end{axis}
\end{tikzpicture}

%% file: figure/ecc_decoder_delay_plot.tex
\begin{tikzpicture}
\begin{axis}[
    ecc axis,
    xmin=0.75, xmax=4.25,
    ymin=0.25, ymax=1.35,
    xtick={1,2,3,4},
    xticklabels={128,256,512,1024},
    xlabel={$k$},
    ylabel={Dec. Delay (ns)},
    xlabel style={yshift=2pt},
    ylabel style={xshift=-6pt, yshift=-2pt},
]
\addplot[ecc t1] coordinates {(1,0.3472) (2,0.3926) (3,0.4439) (4,0.5194)};
\addplot[ecc t2] coordinates {(1,0.5781) (2,0.6166) (3,0.7086) (4,0.7732)};
\addplot[ecc t3] coordinates {(1,0.7883) (2,0.8381) (3,0.9489) (4,1.0092)};
\addplot[ecc t4] coordinates {(1,1.0299) (2,1.0606) (3,1.1801) (4,1.2540)};
\end{axis}
\end{tikzpicture}

%% file: figure/ecc_encoder_delay_plot.tex
\begin{tikzpicture}
\begin{axis}[
    ecc axis,
    xmin=0.75, xmax=4.25,
    ymin=0.15, ymax=0.32,
    xtick={1,2,3,4},
    xticklabels={128,256,512,1024},
    xlabel={$k$},
    ylabel={Enc. Delay (ns)},
    xlabel style={yshift=2pt},
    ylabel style={xshift=-6pt, yshift=-2pt},
]
\addplot[ecc t1] coordinates {(1,0.1877) (2,0.2192) (3,0.2493) (4,0.2803)};
\addplot[ecc t2] coordinates {(1,0.1909) (2,0.2209) (3,0.2512) (4,0.2866)};
\addplot[ecc t3] coordinates {(1,0.1954) (2,0.2305) (3,0.2546) (4,0.2955)};
\addplot[ecc t4] coordinates {(1,0.1930) (2,0.2257) (3,0.2601) (4,0.2936)};
\end{axis}
\end{tikzpicture}

%% file: figure/ecc_decoder_power_plot.tex
\begin{tikzpicture}
\begin{axis}[
    ecc axis,
    height=\eccbaraxisheight,
    xmin=0.45, xmax=17.45,
    ymin=0, ymax=90,
    xtick={1,2,3,4,5.3,6.3,7.3,8.3,9.6,10.6,11.6,12.6,13.9,14.9,15.9,16.9},
    xticklabels={1,2,3,4,1,2,3,4,1,2,3,4,1,2,3,4},
    xlabel={Correction Capability $t$},
    ylabel={Dec. Energy (fJ/bit)},
    clip=false,
    xlabel style={yshift=2pt},
    ylabel style={xshift=-4pt, yshift=-2pt},
]
\newcommand{\eccdecodersegment}[5]{%
    \draw[draw=black!85, line width=0.25pt, fill=#4, fill opacity=#5]
        (axis cs:{#1-0.3},#2) rectangle (axis cs:{#1+0.3},#3);%
}
\eccdecodersegment{1}{0.00}{6.03}{ecctone}{0.18}
\eccdecodersegment{1}{6.03}{6.39}{ecctone}{0.38}
\eccdecodersegment{2}{0.00}{6.67}{eccttwo}{0.18}
\eccdecodersegment{2}{6.67}{20.00}{eccttwo}{0.38}
\eccdecodersegment{2}{20.00}{22.34}{eccttwo}{0.58}
\eccdecodersegment{3}{0.00}{8.00}{ecctthree}{0.18}
\eccdecodersegment{3}{8.00}{28.91}{ecctthree}{0.38}
\eccdecodersegment{3}{28.91}{32.82}{ecctthree}{0.58}
\eccdecodersegment{3}{32.82}{35.79}{ecctthree}{0.78}
\eccdecodersegment{4}{0.00}{11.06}{ecctfour}{0.18}
\eccdecodersegment{4}{11.06}{44.53}{ecctfour}{0.38}
\eccdecodersegment{4}{44.53}{52.97}{ecctfour}{0.58}
\eccdecodersegment{4}{52.97}{57.97}{ecctfour}{0.78}
\eccdecodersegment{4}{57.97}{60.78}{ecctfour}{1.00}
\eccdecodersegment{5.3}{0.00}{6.07}{ecctone}{0.18}
\eccdecodersegment{5.3}{6.07}{6.37}{ecctone}{0.38}
\eccdecodersegment{6.3}{0.00}{6.87}{eccttwo}{0.18}
\eccdecodersegment{6.3}{6.87}{22.11}{eccttwo}{0.38}
\eccdecodersegment{6.3}{22.11}{24.30}{eccttwo}{0.58}
\eccdecodersegment{7.3}{0.00}{8.28}{ecctthree}{0.18}
\eccdecodersegment{7.3}{8.28}{30.70}{ecctthree}{0.38}
\eccdecodersegment{7.3}{30.70}{34.53}{ecctthree}{0.58}
\eccdecodersegment{7.3}{34.53}{37.50}{ecctthree}{0.78}
\eccdecodersegment{8.3}{0.00}{10.47}{ecctfour}{0.18}
\eccdecodersegment{8.3}{10.47}{41.72}{ecctfour}{0.38}
\eccdecodersegment{8.3}{41.72}{49.45}{ecctfour}{0.58}
\eccdecodersegment{8.3}{49.45}{54.06}{ecctfour}{0.78}
\eccdecodersegment{8.3}{54.06}{56.56}{ecctfour}{1.00}
\eccdecodersegment{9.6}{0.00}{6.09}{ecctone}{0.18}
\eccdecodersegment{9.6}{6.09}{6.29}{ecctone}{0.38}
\eccdecodersegment{10.6}{0.00}{6.60}{eccttwo}{0.18}
\eccdecodersegment{10.6}{6.60}{28.67}{eccttwo}{0.38}
\eccdecodersegment{10.6}{28.67}{30.82}{eccttwo}{0.58}
\eccdecodersegment{11.6}{0.00}{8.52}{ecctthree}{0.18}
\eccdecodersegment{11.6}{8.52}{40.63}{ecctthree}{0.38}
\eccdecodersegment{11.6}{40.63}{48.05}{ecctthree}{0.58}
\eccdecodersegment{11.6}{48.05}{51.57}{ecctthree}{0.78}
\eccdecodersegment{12.6}{0.00}{11.52}{ecctfour}{0.18}
\eccdecodersegment{12.6}{11.52}{56.25}{ecctfour}{0.38}
\eccdecodersegment{12.6}{56.25}{66.41}{ecctfour}{0.58}
\eccdecodersegment{12.6}{66.41}{74.22}{ecctfour}{0.78}
\eccdecodersegment{12.6}{74.22}{77.74}{ecctfour}{1.00}
\eccdecodersegment{13.9}{0.00}{6.33}{ecctone}{0.18}
\eccdecodersegment{13.9}{6.33}{6.43}{ecctone}{0.38}
\eccdecodersegment{14.9}{0.00}{6.78}{eccttwo}{0.18}
\eccdecodersegment{14.9}{6.78}{29.89}{eccttwo}{0.38}
\eccdecodersegment{14.9}{29.89}{32.04}{eccttwo}{0.58}
\eccdecodersegment{15.9}{0.00}{8.55}{ecctthree}{0.18}
\eccdecodersegment{15.9}{8.55}{42.77}{ecctthree}{0.38}
\eccdecodersegment{15.9}{42.77}{50.39}{ecctthree}{0.58}
\eccdecodersegment{15.9}{50.39}{54.10}{ecctthree}{0.78}
\eccdecodersegment{16.9}{0.00}{12.01}{ecctfour}{0.18}
\eccdecodersegment{16.9}{12.01}{61.13}{ecctfour}{0.38}
\eccdecodersegment{16.9}{61.13}{71.09}{ecctfour}{0.58}
\eccdecodersegment{16.9}{71.09}{79.10}{ecctfour}{0.78}
\eccdecodersegment{16.9}{79.10}{83.20}{ecctfour}{1.00}

\draw[black!55, densely dotted, line width=0.45pt] (axis cs:4.65,0) -- (axis cs:4.65,90);
\draw[black!55, densely dotted, line width=0.45pt] (axis cs:8.95,0) -- (axis cs:8.95,90);
\draw[black!55, densely dotted, line width=0.45pt] (axis cs:13.25,0) -- (axis cs:13.25,90);

\node[font=\eccplotfont, anchor=south] at (axis description cs:0.11,1.02) {$k=128$};
\node[font=\eccplotfont, anchor=south] at (axis description cs:0.37,1.02) {$k=256$};
\node[font=\eccplotfont, anchor=south] at (axis description cs:0.63,1.02) {$k=512$};
\node[font=\eccplotfont, anchor=south] at (axis description cs:0.89,1.02) {$k=1024$};
\node[font=\eccplotfont, anchor=north west, fill=white, fill opacity=0.82, text opacity=1, inner sep=0.8pt, align=left] at (axis description cs:0.02,0.97) {Darker bars: more injected errors~$i$};
\end{axis}
\end{tikzpicture}

%% file: figure/workload_selection_map_balanced.tex
\begingroup
\newcommand{\eccmapcellfont}{\small}
\newcommand{\eccmaplabelfont}{\small}
\resizebox{\linewidth}{!}{%
\begin{tikzpicture}[x=0.98cm,y=0.62cm]
\tikzset{eccmapid/.style={circle, draw=black, line width=0.28pt, fill=white, fill opacity=0.76, text opacity=1, inner sep=0pt, minimum size=1.15em, font=\eccmapcellfont}}
\node[anchor=south, font=\eccmaplabelfont] at (0.50,0.10) {512};
\node[anchor=south, font=\eccmaplabelfont] at (1.50,0.10) {1k};
\node[anchor=south, font=\eccmaplabelfont] at (2.50,0.10) {2k};
\node[anchor=south, font=\eccmaplabelfont] at (3.50,0.10) {4k};
\node[anchor=south, font=\eccmaplabelfont] at (4.50,0.10) {8k};
\node[anchor=south, font=\eccmaplabelfont] at (5.50,0.10) {16k};
\node[anchor=east, font=\eccmaplabelfont] at (-0.10,-0.50) {0.02};
\node[anchor=east, font=\eccmaplabelfont] at (-0.10,-1.50) {0.05};
\node[anchor=east, font=\eccmaplabelfont] at (-0.10,-2.50) {0.1};
\node[anchor=east, font=\eccmaplabelfont] at (-0.10,-3.50) {0.2};
\node[anchor=east, font=\eccmaplabelfont] at (-0.10,-4.50) {0.35};
\node[anchor=east, font=\eccmaplabelfont] at (-0.10,-5.50) {0.5};
\node[anchor=east, font=\eccmaplabelfont] at (-0.10,-6.50) {0.7};
\fill[ecctfour!65] (0,0) rectangle ++(1,-1);
\fill[ecctfour!65] (1,0) rectangle ++(1,-1);
\fill[ecctfour!65] (2,0) rectangle ++(1,-1);
\fill[ecctfour!65] (3,0) rectangle ++(1,-1);
\fill[ecctthree!65] (4,0) rectangle ++(1,-1);
\fill[ecctthree!65] (5,0) rectangle ++(1,-1);
\fill[ecctfour!65] (0,-1) rectangle ++(1,-1);
\fill[ecctfour!65] (1,-1) rectangle ++(1,-1);
\fill[ecctfour!65] (2,-1) rectangle ++(1,-1);
\fill[ecctthree!65] (3,-1) rectangle ++(1,-1);
\fill[ecctthree!65] (4,-1) rectangle ++(1,-1);
\fill[ecctthree!65] (5,-1) rectangle ++(1,-1);
\fill[ecctfour!65] (0,-2) rectangle ++(1,-1);
\fill[ecctfour!65] (1,-2) rectangle ++(1,-1);
\fill[ecctthree!65] (2,-2) rectangle ++(1,-1);
\fill[ecctthree!65] (3,-2) rectangle ++(1,-1);
\fill[ecctthree!65] (4,-2) rectangle ++(1,-1);
\fill[ecctthree!65] (5,-2) rectangle ++(1,-1);
\fill[ecctfour!65] (0,-3) rectangle ++(1,-1);
\fill[ecctthree!65] (1,-3) rectangle ++(1,-1);
\fill[ecctthree!65] (2,-3) rectangle ++(1,-1);
\fill[ecctthree!65] (3,-3) rectangle ++(1,-1);
\fill[ecctthree!65] (4,-3) rectangle ++(1,-1);
\fill[eccttwo!78] (5,-3) rectangle ++(1,-1);
\fill[ecctthree!65] (0,-4) rectangle ++(1,-1);
\fill[ecctthree!65] (1,-4) rectangle ++(1,-1);
\fill[ecctthree!65] (2,-4) rectangle ++(1,-1);
\fill[ecctthree!65] (3,-4) rectangle ++(1,-1);
\fill[eccttwo!78] (4,-4) rectangle ++(1,-1);
\fill[ecctone!65] (5,-4) rectangle ++(1,-1);
\fill[ecctthree!65] (0,-5) rectangle ++(1,-1);
\fill[ecctthree!65] (1,-5) rectangle ++(1,-1);
\fill[ecctthree!65] (2,-5) rectangle ++(1,-1);
\fill[ecctthree!65] (3,-5) rectangle ++(1,-1);
\fill[eccttwo!78] (4,-5) rectangle ++(1,-1);
\fill[ecctone!65] (5,-5) rectangle ++(1,-1);
\fill[ecctthree!65] (0,-6) rectangle ++(1,-1);
\fill[ecctthree!65] (1,-6) rectangle ++(1,-1);
\fill[ecctthree!65] (2,-6) rectangle ++(1,-1);
\fill[eccttwo!78] (3,-6) rectangle ++(1,-1);
\fill[ecctone!65] (4,-6) rectangle ++(1,-1);
\fill[ecctone!65] (5,-6) rectangle ++(1,-1);
\draw[white, line width=0.32pt] (4,0) -- (4,-1);
\draw[white, line width=0.32pt] (3,-1) -- (3,-2);
\draw[white, line width=0.32pt] (2,-2) -- (2,-3);
\draw[white, line width=0.32pt] (1,-3) -- (1,-4);
\draw[white, line width=0.32pt] (5,-3) -- (5,-4);
\draw[white, line width=0.32pt] (4,-4) -- (4,-5);
\draw[white, line width=0.32pt] (5,-4) -- (5,-5);
\draw[white, line width=0.32pt] (4,-5) -- (4,-6);
\draw[white, line width=0.32pt] (5,-5) -- (5,-6);
\draw[white, line width=0.32pt] (3,-6) -- (3,-7);
\draw[white, line width=0.32pt] (4,-6) -- (4,-7);
\draw[white, line width=0.32pt] (3,-1) -- (4,-1);
\draw[white, line width=0.32pt] (2,-2) -- (3,-2);
\draw[white, line width=0.32pt] (1,-3) -- (2,-3);
\draw[white, line width=0.32pt] (5,-3) -- (6,-3);
\draw[white, line width=0.32pt] (0,-4) -- (1,-4);
\draw[white, line width=0.32pt] (4,-4) -- (5,-4);
\draw[white, line width=0.32pt] (5,-4) -- (6,-4);
\draw[white, line width=0.32pt] (3,-6) -- (4,-6);
\draw[white, line width=0.32pt] (4,-6) -- (5,-6);
\node[eccmapid] at (1.50,-1.50) {4};
\node[eccmapid] at (3.00,-3.50) {3};
\node[eccmapid] at (4.50,-5.00) {2};
\node[eccmapid] at (5.50,-5.50) {1};
\draw[black, line width=0.34pt] (0,0) rectangle (6,-7);
\node[anchor=north, font=\eccmaplabelfont] at (3,-7.12) {$BW$ (bits/cycle)};
\node[rotate=90, anchor=south, font=\eccmaplabelfont] at (-0.74,-3.50) {Base Activity $\alpha$};
\end{tikzpicture}%
}
\endgroup

%% file: figure/workload_selection_map_read_heavy.tex
\begingroup
\newcommand{\eccmapcellfont}{\small}
\newcommand{\eccmaplabelfont}{\small}
\resizebox{\linewidth}{!}{%
\begin{tikzpicture}[x=0.98cm,y=0.62cm]
\tikzset{eccmapid/.style={circle, draw=black, line width=0.28pt, fill=white, fill opacity=0.76, text opacity=1, inner sep=0pt, minimum size=1.15em, font=\eccmapcellfont}}
\node[anchor=south, font=\eccmaplabelfont] at (0.50,0.10) {512};
\node[anchor=south, font=\eccmaplabelfont] at (1.50,0.10) {1k};
\node[anchor=south, font=\eccmaplabelfont] at (2.50,0.10) {2k};
\node[anchor=south, font=\eccmaplabelfont] at (3.50,0.10) {4k};
\node[anchor=south, font=\eccmaplabelfont] at (4.50,0.10) {8k};
\node[anchor=south, font=\eccmaplabelfont] at (5.50,0.10) {16k};
\node[anchor=east, font=\eccmaplabelfont] at (-0.10,-0.50) {0.02};
\node[anchor=east, font=\eccmaplabelfont] at (-0.10,-1.50) {0.05};
\node[anchor=east, font=\eccmaplabelfont] at (-0.10,-2.50) {0.1};
\node[anchor=east, font=\eccmaplabelfont] at (-0.10,-3.50) {0.2};
\node[anchor=east, font=\eccmaplabelfont] at (-0.10,-4.50) {0.35};
\node[anchor=east, font=\eccmaplabelfont] at (-0.10,-5.50) {0.5};
\node[anchor=east, font=\eccmaplabelfont] at (-0.10,-6.50) {0.7};
\fill[ecctfour!65] (0,0) rectangle ++(1,-1);
\fill[ecctfour!65] (1,0) rectangle ++(1,-1);
\fill[ecctfour!65] (2,0) rectangle ++(1,-1);
\fill[ecctfour!65] (3,0) rectangle ++(1,-1);
\fill[ecctthree!65] (4,0) rectangle ++(1,-1);
\fill[ecctthree!65] (5,0) rectangle ++(1,-1);
\fill[ecctfour!65] (0,-1) rectangle ++(1,-1);
\fill[ecctfour!65] (1,-1) rectangle ++(1,-1);
\fill[ecctfour!65] (2,-1) rectangle ++(1,-1);
\fill[ecctthree!65] (3,-1) rectangle ++(1,-1);
\fill[ecctthree!65] (4,-1) rectangle ++(1,-1);
\fill[ecctthree!65] (5,-1) rectangle ++(1,-1);
\fill[ecctfour!65] (0,-2) rectangle ++(1,-1);
\fill[ecctfour!65] (1,-2) rectangle ++(1,-1);
\fill[ecctthree!65] (2,-2) rectangle ++(1,-1);
\fill[ecctthree!65] (3,-2) rectangle ++(1,-1);
\fill[ecctthree!65] (4,-2) rectangle ++(1,-1);
\fill[ecctthree!65] (5,-2) rectangle ++(1,-1);
\fill[ecctfour!65] (0,-3) rectangle ++(1,-1);
\fill[ecctthree!65] (1,-3) rectangle ++(1,-1);
\fill[ecctthree!65] (2,-3) rectangle ++(1,-1);
\fill[ecctthree!65] (3,-3) rectangle ++(1,-1);
\fill[ecctthree!65] (4,-3) rectangle ++(1,-1);
\fill[eccttwo!78] (5,-3) rectangle ++(1,-1);
\fill[ecctthree!65] (0,-4) rectangle ++(1,-1);
\fill[ecctthree!65] (1,-4) rectangle ++(1,-1);
\fill[ecctthree!65] (2,-4) rectangle ++(1,-1);
\fill[ecctthree!65] (3,-4) rectangle ++(1,-1);
\fill[eccttwo!78] (4,-4) rectangle ++(1,-1);
\fill[eccttwo!78] (5,-4) rectangle ++(1,-1);
\fill[ecctthree!65] (0,-5) rectangle ++(1,-1);
\fill[ecctthree!65] (1,-5) rectangle ++(1,-1);
\fill[ecctthree!65] (2,-5) rectangle ++(1,-1);
\fill[ecctthree!65] (3,-5) rectangle ++(1,-1);
\fill[eccttwo!78] (4,-5) rectangle ++(1,-1);
\fill[eccttwo!78] (5,-5) rectangle ++(1,-1);
\fill[ecctthree!65] (0,-6) rectangle ++(1,-1);
\fill[ecctthree!65] (1,-6) rectangle ++(1,-1);
\fill[ecctthree!65] (2,-6) rectangle ++(1,-1);
\fill[eccttwo!78] (3,-6) rectangle ++(1,-1);
\fill[eccttwo!78] (4,-6) rectangle ++(1,-1);
\fill[ecctone!65] (5,-6) rectangle ++(1,-1);
\draw[white, line width=0.32pt] (4,0) -- (4,-1);
\draw[white, line width=0.32pt] (3,-1) -- (3,-2);
\draw[white, line width=0.32pt] (2,-2) -- (2,-3);
\draw[white, line width=0.32pt] (1,-3) -- (1,-4);
\draw[white, line width=0.32pt] (5,-3) -- (5,-4);
\draw[white, line width=0.32pt] (4,-4) -- (4,-5);
\draw[white, line width=0.32pt] (4,-5) -- (4,-6);
\draw[white, line width=0.32pt] (3,-6) -- (3,-7);
\draw[white, line width=0.32pt] (5,-6) -- (5,-7);
\draw[white, line width=0.32pt] (3,-1) -- (4,-1);
\draw[white, line width=0.32pt] (2,-2) -- (3,-2);
\draw[white, line width=0.32pt] (1,-3) -- (2,-3);
\draw[white, line width=0.32pt] (5,-3) -- (6,-3);
\draw[white, line width=0.32pt] (0,-4) -- (1,-4);
\draw[white, line width=0.32pt] (4,-4) -- (5,-4);
\draw[white, line width=0.32pt] (3,-6) -- (4,-6);
\draw[white, line width=0.32pt] (5,-6) -- (6,-6);
\node[eccmapid] at (1.50,-1.50) {4};
\node[eccmapid] at (3.00,-3.50) {3};
\node[eccmapid] at (5.00,-5.00) {2};
\node[eccmapid] at (5.50,-6.50) {1};
\draw[black, line width=0.34pt] (0,0) rectangle (6,-7);
\node[anchor=north, font=\eccmaplabelfont] at (3,-7.12) {$BW$ (bits/cycle)};
\node[rotate=90, anchor=south, font=\eccmaplabelfont] at (-0.74,-3.50) {Base Activity $\alpha$};
\end{tikzpicture}%
}
\endgroup

%% file: figure/workload_selection_map_write_heavy.tex
\begingroup
\newcommand{\eccmapcellfont}{\small}
\newcommand{\eccmaplabelfont}{\small}
\resizebox{\linewidth}{!}{%
\begin{tikzpicture}[x=0.98cm,y=0.62cm]
\tikzset{eccmapid/.style={circle, draw=black, line width=0.28pt, fill=white, fill opacity=0.76, text opacity=1, inner sep=0pt, minimum size=1.15em, font=\eccmapcellfont}}
\node[anchor=south, font=\eccmaplabelfont] at (0.50,0.10) {512};
\node[anchor=south, font=\eccmaplabelfont] at (1.50,0.10) {1k};
\node[anchor=south, font=\eccmaplabelfont] at (2.50,0.10) {2k};
\node[anchor=south, font=\eccmaplabelfont] at (3.50,0.10) {4k};
\node[anchor=south, font=\eccmaplabelfont] at (4.50,0.10) {8k};
\node[anchor=south, font=\eccmaplabelfont] at (5.50,0.10) {16k};
\node[anchor=east, font=\eccmaplabelfont] at (-0.10,-0.50) {0.02};
\node[anchor=east, font=\eccmaplabelfont] at (-0.10,-1.50) {0.05};
\node[anchor=east, font=\eccmaplabelfont] at (-0.10,-2.50) {0.1};
\node[anchor=east, font=\eccmaplabelfont] at (-0.10,-3.50) {0.2};
\node[anchor=east, font=\eccmaplabelfont] at (-0.10,-4.50) {0.35};
\node[anchor=east, font=\eccmaplabelfont] at (-0.10,-5.50) {0.5};
\node[anchor=east, font=\eccmaplabelfont] at (-0.10,-6.50) {0.7};
\fill[ecctfour!65] (0,0) rectangle ++(1,-1);
\fill[ecctfour!65] (1,0) rectangle ++(1,-1);
\fill[ecctfour!65] (2,0) rectangle ++(1,-1);
\fill[ecctfour!65] (3,0) rectangle ++(1,-1);
\fill[ecctfour!65] (4,0) rectangle ++(1,-1);
\fill[ecctfour!65] (5,0) rectangle ++(1,-1);
\fill[ecctfour!65] (0,-1) rectangle ++(1,-1);
\fill[ecctfour!65] (1,-1) rectangle ++(1,-1);
\fill[ecctfour!65] (2,-1) rectangle ++(1,-1);
\fill[ecctfour!65] (3,-1) rectangle ++(1,-1);
\fill[ecctfour!65] (4,-1) rectangle ++(1,-1);
\fill[ecctthree!65] (5,-1) rectangle ++(1,-1);
\fill[ecctfour!65] (0,-2) rectangle ++(1,-1);
\fill[ecctfour!65] (1,-2) rectangle ++(1,-1);
\fill[ecctfour!65] (2,-2) rectangle ++(1,-1);
\fill[ecctfour!65] (3,-2) rectangle ++(1,-1);
\fill[ecctthree!65] (4,-2) rectangle ++(1,-1);
\fill[ecctthree!65] (5,-2) rectangle ++(1,-1);
\fill[ecctfour!65] (0,-3) rectangle ++(1,-1);
\fill[ecctfour!65] (1,-3) rectangle ++(1,-1);
\fill[ecctfour!65] (2,-3) rectangle ++(1,-1);
\fill[ecctthree!65] (3,-3) rectangle ++(1,-1);
\fill[ecctthree!65] (4,-3) rectangle ++(1,-1);
\fill[ecctthree!65] (5,-3) rectangle ++(1,-1);
\fill[ecctfour!65] (0,-4) rectangle ++(1,-1);
\fill[ecctfour!65] (1,-4) rectangle ++(1,-1);
\fill[ecctthree!65] (2,-4) rectangle ++(1,-1);
\fill[ecctthree!65] (3,-4) rectangle ++(1,-1);
\fill[ecctthree!65] (4,-4) rectangle ++(1,-1);
\fill[eccttwo!78] (5,-4) rectangle ++(1,-1);
\fill[ecctfour!65] (0,-5) rectangle ++(1,-1);
\fill[ecctthree!65] (1,-5) rectangle ++(1,-1);
\fill[ecctthree!65] (2,-5) rectangle ++(1,-1);
\fill[ecctthree!65] (3,-5) rectangle ++(1,-1);
\fill[ecctthree!65] (4,-5) rectangle ++(1,-1);
\fill[eccttwo!78] (5,-5) rectangle ++(1,-1);
\fill[ecctfour!65] (0,-6) rectangle ++(1,-1);
\fill[ecctthree!65] (1,-6) rectangle ++(1,-1);
\fill[ecctthree!65] (2,-6) rectangle ++(1,-1);
\fill[ecctthree!65] (3,-6) rectangle ++(1,-1);
\fill[eccttwo!78] (4,-6) rectangle ++(1,-1);
\fill[ecctone!65] (5,-6) rectangle ++(1,-1);
\draw[white, line width=0.32pt] (5,-1) -- (5,-2);
\draw[white, line width=0.32pt] (4,-2) -- (4,-3);
\draw[white, line width=0.32pt] (3,-3) -- (3,-4);
\draw[white, line width=0.32pt] (2,-4) -- (2,-5);
\draw[white, line width=0.32pt] (5,-4) -- (5,-5);
\draw[white, line width=0.32pt] (1,-5) -- (1,-6);
\draw[white, line width=0.32pt] (5,-5) -- (5,-6);
\draw[white, line width=0.32pt] (1,-6) -- (1,-7);
\draw[white, line width=0.32pt] (4,-6) -- (4,-7);
\draw[white, line width=0.32pt] (5,-6) -- (5,-7);
\draw[white, line width=0.32pt] (5,-1) -- (6,-1);
\draw[white, line width=0.32pt] (4,-2) -- (5,-2);
\draw[white, line width=0.32pt] (3,-3) -- (4,-3);
\draw[white, line width=0.32pt] (2,-4) -- (3,-4);
\draw[white, line width=0.32pt] (5,-4) -- (6,-4);
\draw[white, line width=0.32pt] (1,-5) -- (2,-5);
\draw[white, line width=0.32pt] (4,-6) -- (5,-6);
\draw[white, line width=0.32pt] (5,-6) -- (6,-6);
\node[eccmapid] at (2.00,-2.50) {4};
\node[eccmapid] at (4.00,-4.50) {3};
\node[eccmapid] at (5.50,-5.00) {2};
\node[eccmapid] at (5.50,-6.50) {1};
\draw[black, line width=0.34pt] (0,0) rectangle (6,-7);
\node[anchor=north, font=\eccmaplabelfont] at (3,-7.12) {$BW$ (bits/cycle)};
\node[rotate=90, anchor=south, font=\eccmaplabelfont] at (-0.74,-3.50) {Base Activity $\alpha$};
\end{tikzpicture}%
}
\endgroup

%% file: figure/workload_power_sweep_balanced.tex
\begingroup
\begin{tikzpicture}
\begin{axis}[
    ecc axis,
    height=0.58\linewidth,
    xlabel={Workload intensity $BW\times\alpha$},
    ylabel={Total power (mW)},
    xmin=0, xmax=12000,
    ymin=0, ymax=690,
    xtick={0,2000,4000,6000,8000,10000,12000},
    xticklabels={0,2k,4k,6k,8k,10k,12k},
    scaled x ticks=false,
    ytick={0,100,200,300,400,500,600},
    xmajorgrids=true,
    legend columns=3,
    legend image code/.code={%
        \draw[mark repeat=2, mark phase=2, #1]
            plot coordinates {(0cm,0cm) (0.16cm,0cm) (0.32cm,0cm)};%
    },
    legend style={
        draw=black,
        fill=white,
        fill opacity=0.95,
        text opacity=1,
        font=\eccplotfont,
        cells={anchor=west},
        inner sep=1.4pt,
        at={(0.02,0.98)},
        anchor=north west
    },
]
\addplot+[black!65, line width=0.65pt, mark=*, mark size=1.25pt, mark options={fill=black!65}]
coordinates {
    (10.24,367.796) (20.48,367.968) (25.60,368.054)
    (40.96,368.312) (51.20,368.484) (51.20,368.484)
    (81.92,369.000) (102.40,369.345) (102.40,369.345)
    (102.40,369.345) (163.84,370.377) (179.20,370.635)
    (204.80,371.065) (204.80,371.065) (204.80,371.065)
    (256.00,371.925) (327.68,373.129) (358.40,373.645)
    (358.40,373.645) (409.60,374.506) (409.60,374.506)
    (409.60,374.506) (512.00,376.226) (716.80,379.666)
    (716.80,379.666) (819.20,381.387) (819.20,381.387)
    (819.20,381.387) (1024.00,384.827) (1433.60,391.709)
    (1433.60,391.709) (1638.40,395.149) (1638.40,395.149)
    (2048.00,402.031) (2867.20,415.793) (2867.20,415.793)
    (3276.80,422.674) (4096.00,436.437) (5734.40,463.962)
    (5734.40,463.962) (8192.00,505.250) (11468.80,560.300)
};
\addlegendentry{No ECC}
\addplot+[ecctone, line width=0.65pt, mark=square*, mark size=1.25pt, mark options={fill=ecctone}]
coordinates {
    (10.24,74.472) (20.48,74.644) (25.60,74.854)
    (40.96,75.071) (51.20,75.490) (51.20,75.284)
    (81.92,75.924) (102.40,76.763) (102.40,76.350)
    (102.40,76.350) (163.84,77.630) (179.20,78.672)
    (204.80,78.483) (204.80,78.483) (204.80,78.483)
    (256.00,80.581) (327.68,81.042) (358.40,83.126)
    (358.40,81.682) (409.60,82.749) (409.60,82.749)
    (409.60,82.749) (512.00,84.881) (716.80,89.147)
    (716.80,89.147) (819.20,91.280) (819.20,91.280)
    (819.20,91.280) (1024.00,95.545) (1433.60,104.077)
    (1433.60,104.077) (1638.40,108.342) (1638.40,108.342)
    (2048.00,116.873) (2867.20,133.936) (2867.20,133.936)
    (3276.80,142.467) (4096.00,159.529) (5734.40,193.654)
    (5734.40,193.654) (8192.00,244.841) (11468.80,313.090)
};
\addlegendentry{ECC$(1035,1024,1)$}
\addplot+[ecctthree!75!black, line width=0.75pt, mark=diamond*, mark size=1.8pt, mark options={fill=ecctthree!75!black}]
coordinates {
    (10.24,19.048) (20.48,19.459) (25.60,19.665)
    (40.96,20.283) (51.20,20.694) (51.20,20.694)
    (81.92,21.929) (102.40,22.752) (102.40,22.752)
    (102.40,22.752) (163.84,24.630) (179.20,25.086)
    (204.80,25.845) (204.80,25.845) (204.80,25.845)
    (256.00,27.364) (327.68,29.491) (358.40,30.402)
    (358.40,30.402) (409.60,31.921) (409.60,31.921)
    (409.60,31.921) (512.00,34.960) (716.80,41.036)
    (716.80,41.036) (819.20,44.074) (819.20,44.074)
    (819.20,44.074) (1024.00,50.150) (1433.60,62.303)
    (1433.60,62.303) (1638.40,68.379) (1638.40,68.379)
    (2048.00,80.532) (2867.20,101.587) (2867.20,101.587)
    (3276.80,111.967) (4096.00,132.726) (5734.40,174.241)
    (5734.40,174.241) (8192.00,227.340) (11468.80,298.139)
};
\addlegendentry{Selected ECC}
\draw[<-, >=latex, black!70, line width=0.55pt]
    (axis cs:430,35.048) -- (axis cs:430,360.796);
\node[font=\eccplotfont, anchor=west, fill=white, fill opacity=0.86, text opacity=1,
      inner sep=0.8pt] at (axis cs:360,193.4) {$94.8\%$};
\draw[<-, >=latex, black!70, line width=0.55pt]
    (axis cs:11250,289.139) -- (axis cs:11250,558.300);
\node[font=\eccplotfont, anchor=east, fill=white, fill opacity=0.86, text opacity=1,
      inner sep=0.8pt] at (axis cs:11680,429.2) {$46.8\%$};
\end{axis}
\end{tikzpicture}
\endgroup

%% file: IEEEabrv.bib
@STRING{IEEE_J_CAD        = "{IEEE} Trans. Comput.-Aided Design Integr. Circuits Syst."}

@STRING{IEEE_J_JLT        = "J. Lightw. Technol."}

@STRING{IEEE_J_JSSC       = "{IEEE} J. Solid-State Circuits"}

@STRING{IEEE_J_VLSI       = "{IEEE} Trans. {VLSI} Syst."}


%% file: mybib.bib
@book{meinerzhagen2018gain,
  title={Gain-cell Embedded {DRAMs} for Low-power {VLSI} Systems-on-chip},
  author={Meinerzhagen, Pascal and Teman, Adam and Giterman, Robert and Edri, Noa and Burg, Andreas and Fish, Alexander},
  year={2018},
  publisher={Springer}
}

@article{chun2011667,
  title={A 667 {MHz} logic-compatible embedded {DRAM} featuring an asymmetric {2T} gain cell for high speed on-die caches},
  author={Chun, Ki Chul and Jain, Pulkit and Kim, Tae-Ho and Kim, Chris H},
  journal=IEEE_J_JSSC,
  volume={47},
  number={2},
  pages={547--559},
  year={2012},
  month={Feb.}
}

@article{chun20113t,
  title={A {3T} gain cell embedded {DRAM} utilizing preferential boosting for high density and low power on-die caches},
  author={Chun, Ki Chul and Jain, Pulkit and Lee, Jung Hwa and Kim, Chris H},
  journal=IEEE_J_JSSC,
  volume={46},
  number={6},
  pages={1495--1505},
  year={2011},
  month={Jun.}
}

@article{nguyen2024mcaimem,
  title={{MCAIMem}: A mixed {SRAM} and {eDRAM} cell for area and energy-efficient on-chip {AI} memory},
  author={Nguyen, Duy-Thanh and Bhattacharjee, Abhiroop and Moitra, Abhishek and Panda, Priyadarshini},
  journal=IEEE_J_VLSI,
  volume={32},
  number={11},
  pages={2023-2036},
  year={2024},
  month={Nov.}
}

@inproceedings{wilkerson2010reducing,
  title={Reducing cache power with low-cost, multi-bit error-correcting codes},
  author={Wilkerson, Chris and Alameldeen, Alaa R and Chishti, Zeshan and Wu, Wei and Somasekhar, Dinesh and Lu, Shih-Lien},
  booktitle={ISCA},
  pages={83--93},
  year={2010}
}

@article{joshi2026modeling,
  title={Modeling energy and delay for {ECC}-enabled cache architectures},
  author={Joshi, Dinesh and Bagchi, Aritra and Agarwal, Ayushi and Modi, Garima and Srivastava, Neha and Roy, Sourav and Panda, Preeti Ranjan},
  journal=IEEE_J_CAD,
  year={Early access, 2026},
  doi={10.1109/TCAD.2026.3665117}
}

@inproceedings{chen2021care,
  title={{CARE}: Coordinated augmentation for elastic resilience on {DRAM} errors in data centers},
  author={Chen, Jian and Jiang, Xiaowei and Zhang, Ying and Liu, Liyin and Xu, Huifeng and Liu, Qiang},
  booktitle={HPCA},
  pages={533--544},
  year={2021}
}

@inproceedings{lee2022stealth,
  title={Stealth {ECC}: A data-width aware adaptive {ECC} scheme for {DRAM} error resilience},
  author={Lee, Young Seo and Koo, Gunjae and Gong, Young-Ho and Chung, Sung Woo},
  booktitle={DATE},
  pages={382--387},
  year={2022}
}

@inproceedings{lee2025shift,
  title={{SHIFT} {ECC}: A value converting {HBM} {ECC} approach for refresh energy efficient integer quantized {DNN} inference},
  author={Lee, Jae Yoon and Lee, Young Seo and Gong, Young-Ho and Kim, Seon Wook and Chung, Sung Woo},
  booktitle={ISLPED},
  pages={1--7},
  year={2025}
}

@article{harel202416,
  title={A 16-k{B} 65-nm {GC-eDRAM} macro with internal bias voltage generation providing over 100-$\mu$s retention time},
  author={Harel, Odem and Yigit, Andac and Feifel, Eliana and Giterman, Robert and Burg, Andreas and Teman, Adam},
  journal=IEEE_J_JSSC,
  volume={60},
  number={6},
  pages={2239-2248},
  year={2024},
  month={Jun.}
}

@article{fougstedt2019energy,
  title={Energy-efficient high-throughput {VLSI} architectures for product-like codes},
  author={Fougstedt, Christoffer and Larsson-Edefors, Per},
  journal=IEEE_J_JLT,
  volume={37},
  number={2},
  pages={477--485},
  year={2019},
  month={Jan.}
}

@article{li2025gainsight,
  title={{GainSight}: Application-guided profiling for composing heterogeneous on-chip memories in {AI} hardware accelerators},
  author={Li, Peijing and others},
  journal={arXiv preprint arXiv:2504.14866},
  year={2025}
}

@article{wang2025opengcram,
  title={{OpenGCRAM}: An open-source gain cell compiler enabling design-space exploration for {AI} workloads},
  author={Wang, Xinxin and others},
  journal={arXiv preprint arXiv:2507.10849},
  year={2025}
}

@article{choi2018decoder,
  title={A decoder for short {BCH} codes with high decoding efficiency and low power for emerging memories},
  author={Choi, Sara and Ahn, Hong Keun and Song, Byung Kyu and Kim, Jung Pill and Kang, Seung H and Jung, Seong-Ook},
  journal=IEEE_J_VLSI,
  volume={27},
  number={2},
  pages={387--397},
  year={2019},
  month={Feb.}
}

@inproceedings{rohman2024fast,
  title={Fast energy optimization of on-chip {ECC} memories},
  author={Rohman, Shahriar and Leduc-Primeau, Fran{\c{c}}ois},
  booktitle={SiPS},
  pages={207--212},
  year={2024}
}

@inproceedings{andac25a512,
  title={A 512~kb 0.069~$\mu$m$^2$ logic {3T} {GCRAM} with 27~$\mu$s retention time at 85~$^\circ$C in 16~nm {FinFET}},
  author={Yigit, A. and others},
  booktitle={ESSCIRC},
  pages={293--296},
  year={2025}
}
